\documentclass[proof]{WileyASNA-v1}

\makeatletter
\renewcommand{\itemautorefname}{\@gobble}
\makeatother

\articletype{Conference proceeding}%

\received{01/07/2021}
\revised{28/09/2021}
\accepted{}

\begin{document}

\title{Characterizing X-ray activity cycles of young solar-like stars with solar observations}

\author[1]{Coffaro M.*}

\author[1,2]{Stelzer B.}

\author[2]{Orlando S.}

\authormark{Coffaro M. \textsc{et al}}

\address[1]{\orgdiv{Institut f\"ur Astronomie und Astrophysik T\"ubingen (IAAT)}, \orgname{Eberhard-Karls Universit\"at T\"ubingen}, \orgaddress{ Sand 1, D-72076, \state{T\"ubingen}, \country{Germany}}}

\address[2]{\orgdiv{INAF - Osservatorio Astronomico di Palermo}, \orgaddress{Piazza del Parlamento 1, I-90134, \state{Palermo}, \country{Italy}}}

\corres{*\email{coffaro@astro.uni-tuebingen.de}}


\abstract{
 Throughout an activity cycle, magnetic structures rise to the stellar surface, evolve and decay. Tracing their evolution on a stellar corona allows us to characterize the X-ray cycles. However, directly mapping magnetic structures is feasible only for the Sun, while such structures are spatially unresolved with present-day X-ray instruments on stellar coronae. We present here a new method, implemented by us, that indirectly reproduces the stellar X-ray spectrum and its variability with solar magnetic structures. The technique converts solar corona observations into a format identical to that of stellar X-ray observations and, specifically, \textit{XMM-Newton} spectra. From matching these synthetic spectra with those observed for a star of interest, a fractional surface coverage with solar magnetic structures can be associated to each X-ray observation. We apply this method to two young solar-like stars: $\epsilon$~Eri ($\sim 400$~Myr), the youngest star to display a coronal cycle ($\sim 3$~yr), and Kepler~63 ($\sim 200$~Myr), for which the X-ray monitoring did not reveal a cyclic variability.  We found that even during the cycle minimum a large portion of $\epsilon$ Eri's corona is covered with active structures. Therefore, there is little space for additional magnetic regions during the maximum, explaining the small observed cycle amplitude ($\Delta f \sim 0.12$) in terms of the X-ray luminosity. 
 Kepler~63 displays an even higher coverage with magnetic structure than the corona of $\epsilon$~Eri. This supports the hypothesis that for stars younger than $<400$~Myr the X-ray cycles are inhibited by a massive presence of coronal regions. 
 }

\jnlcitation{\cname{%
\author{Coffaro M.}, 
\author{Stelzer B.}, and 
\author{Orlando S.}} (\cyear{2021}), 
\ctitle{The X-ray activity cycles of young solar-like stars.}, \cjournal{}, \cvol{}.}


\maketitle

\section{Introduction}
The activity cycle of the Sun is defined as the time for the magnetic field to change its configuration: the $\alpha-\Omega$ dynamo, that maintains the magnetic field of the star, twists the magnetic lines from a dipole configuration to a toroidal, and back. 
A direct consequence of this phenomenon is the raise on the surface of magnetic structures, manifested in all layers of the solar atmosphere. Throughout the well-known 11-yr solar cycle, the number of these structures changes periodically, going from a minimum to a maximum surface fraction coverage. 



The most evident structures, visible in the photosphere of the Sun and detectable in optical wavebands, are the sunspots. In the chromosphere, the manifestations of the magnetic spots are the so-called plages, that emit, amongst others, the H\&K lines of Ca\,II. Magnetic structures can also be observed in the corona of the Sun, emitting in the X-rays, known as coronal loops and hosting several structures such as the active regions. 

Characterizing an activity cycle from direct observations of the magnetic structures is feasible only for the Sun. With present-day instruments, it is not possible to spatially resolve these structures on other stars. However, several studies had been carried out in the last decades, enabling the possibility of indirectly studying and characterizing the activity cycles of solar-like stars. 

One of the most well known study was the Mt. Wilson project \citep{1978ApJ...226..379W}. It started in the 1960s and ended in the early 1990s. This project had the aim of monitoring the variation of the Ca\,II H\&K lines emission in a large sample of solar-like stars. The so-called S-index was defined, a parameter that estimates the flux of the Ca\,II emission. The Mt. Wilson project showed that $\sim 60\%$ of main-sequence stars exhibit periodic changes of the H\&K flux, i.e. they have chromospheric cycles with periods up to 20 yr.

Since the underlying cause for cyclic brightness variations in all layers of the atmosphere are magnetic fields, it is expected that the X-ray luminosity (representing the corona) varies in line with the chromospheric Ca\,II emission. 
Up to date, the X-ray satellite \textit{XMM-Newton} revealed the presence of X-ray activity cycles in seven solar-like stars: $\alpha$ Cen A and B ($15$ and $8.8$~yr; \citealt{2012A&A...543A..84R, 2017MNRAS.464.3281W}), 61 Cyg A and B ($7.3$ and $11.3$~yr; \citealt{2006A&A...460..261H, 2012A&A...543A..84R}), HD 81809 ($8$~yr; \citealt{2008A&A...490.1121F, 2017A&A...605A..19O}), $\iota$~Hor ($1.6$~yr; \citealt{2013A&A...553L...6S, 2019A&A...631A..45S}) and $\epsilon$~Eri ($2.95$~yr; \citealt{2020A&A...636A..49C}). All stars were monitored in the chromosphere and the variation of their X-ray fluxes coincides with that of the S-index. 
In this sample, $\iota$~Hor and $\epsilon$~Eri are the two youngest stars with the shortest cycles, $1.6$ and $2.95$~yr respectively. Interestingly they show also the smallest brightness amplitude of their cycles. To further explore the regime of short cycles, we undertook an \textit{XMM-Newton} monitoring campaign for the solar-like star Kepler 63 (Coffaro et al. 2021, submitted to A\&A). With an age of $\sim 200$ Myr \citep{2013ApJ...775...54S} and a photospheric cycle of $1.27$~yr, it is the youngest star monitored in the X-ray band so far with the aim of identifying a coronal cycle\footnote{There is no long-term monitoring of the chromospheric index of Kepler~63. Therefore, no information of its chromospheric cycle is available.}. 

In this article we present a novel method to characterize the X-ray emission of stellar coronae, that in a first rough version had been applied only to HD~81809 \citep{2008A&A...490.1121F, 2017A&A...605A..19O} and we used it for $\epsilon$~Eri and Kepler~63. In a refined version, implemented by us, the method aims to quantitatively investigate the magnetic coronal structures without directly observing them. The technique makes use of the solar magnetic structures, such as active regions (ARs), cores of active regions (COs) and flares (FLs), observed on the Sun  with the satellite \textit{Yohkoh} in the 1990s. The solar X-ray observations are then converted into synthetic spectra as if they were acquired by non-solar X-ray observatories  such as, for instance, \textit{XMM-Newton}. By comparing these synthetic spectra with the actual \textit{XMM-Newton} spectra of $\epsilon$~Eri and Kepler~63, we quantify the type and the percentage of magnetic structures responsible for the X-ray activity of these stars.

In \autoref{sec:sim} we summarize the technical details of our method. In \autoref{sec:results} we present the main results obtained for $\epsilon$~Eri and Kepler~63. Finally, in \autoref{sec:conc} we give our discussions and conclusions.

\section{Simulating X-ray emitting magnetic structures}
\label{sec:sim}
Our method is based on the study "The Sun as an X-ray star", carried out at INAF - Osservatorio Astronomico di Palermo in the early 2000s \citep{2000ApJ...528..524O, 2000ApJ...528..537P,2001ApJ...557..906R, 2001ApJ...560..499O}. Within this study, images of the full Sun collected with \textit{Yohkoh} were used to identify different types of coronal structures (ARs, COs, FLs) and derive, for each of them, the distribution of emission measure versus temperature, EMD.
\begin{figure}[!htbp]
\centering
\includegraphics[width=0.5\textwidth]{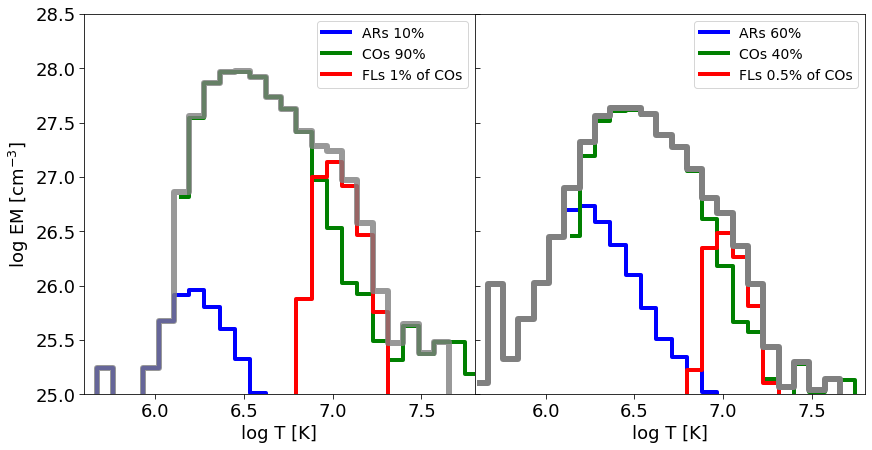}
\caption{Two total EMDs (gray lines) obtained from a combination of ARs, COs and FLs (blue, green and red lines respectively). Each panel shows different coverage fractions with magnetic structures. The percentage of each magnetic structure contributing to the total EMDs is given in the legend of plots.}
\label{fig:arvsco}
\end{figure}

Following these studies, from the solar observations, we derive a representative EMD per unit surface area for each type of solar structures, averaging the EMDs derived for samples of ARs, COs, and FLs over their temporal evolution. Then, we assume a filling factor $f$ for each of the different solar structures on the corona of the star of interest and we derive the total EMD as the sum of the EMDs of the ARs, COs and FLs\footnote{Additionally, the EMDs are scaled to the coronal relative abundances of the star of interest.}. 
When building the total EMD, the filling factor of FLs is tied to the coverage fraction of the COs.  In fact, it is plausible that the FLs, being energetic events, take place within a CO, the brightest coronal structure. Therefore, it is reasonable to assume that among all COs on a stellar corona a certain percentage of them represents flaring events.

In \autoref{fig:arvsco} we show two examples of EMDs, composed of the three types of magnetic structures that contribute at different percentage, as stated in the legend of the plot. Clearly, the total EMD depends strongly on the relative contribution of the magnetic structures.

Finally, the total EMD is translated into an X-ray spectrum with a format identical to that of non-solar X-ray observatories as follows: 

\begin{equation}
\label{eq:1}
C_j = \dfrac{t}{4 \pi d^2} \int dE \dfrac{A(E)M(j,E)}{E}\sum _k P(T_k,E)\text{EMD}
\end{equation}
where $P(T_k, E)$ is the spectrum emitted by an optically thin isothermal plasma, computed using the available spectral codes, at the temperature $T_k$ for photons of energy $E$; $A(E)$ and $M(j,E)$ are the effective area and the response matrix of the instruments (i.e. the probability that a photon of energy $E$ is detected in the $j$th energy channel); $t$ is the (assumed) exposure time of the observation and $d$ is the distance of the simulated star in parsec. These synthetic spectra can be analyzed with the usual tools for spectral analysis, such as \texttt{xspec} \citep{1996ASPC..101...17A}.

To characterize the X-ray activity level of the simulated star, we build a grid of synthetic spectra, each of them derived from an EMD that represents a different combination of magnetic structures. In this way, we can explore different coverage fractions with ARs, COs and FLs on the simulated corona. Then, we analyze these spectra, employing a multi-T spectral model such as the APEC model \citep{2001ApJ...556L..91S}. We obtain thus the best-fitting X-ray luminosity, temperature ($kT$) and emission measure ($EM$). The two latter parameters are used to calculate the EM-averaged coronal temperature as $T_{\rm av} = \dfrac{\sum _i T_i \cdot EM_i}{\sum _i EM_i}$, where $i=1,...,N$ refers to the spectral components used in the fitting. 

To determine uncertainties for a multi-T spectrum, a Monte-Carlo simulation is applied to the grid of synthetic spectra. By introducing statistical noise, we generate N times each spectrum that composes the grid so that we produce N times the same combination of magnetic structures on the stellar corona. Thus, we obtain a set of N best-fitting values for the spectral parameters for each grid point.  The average and the standard deviation of these N values are adopted as the final value and its associated error. 

Finally, we compare the synthetic spectra with the actual observed X-ray spectra of the star of interest, as follows. First, we check if the observed X-ray luminosity and the average temperature derived for the star of interest can be reproduced by the grid of models. Then we move on to a detailed quantitative comparison between the synthetic spectra and the actual observations. We implement a selection criterion that screens the N grids of the synthetic spectra: from each of the N grids the criterion chooses the spectrum whose $kT$ and $EM$ are the closest to the ones derived from the observations of the star, within a confidence level $\sigma$. This way, we identify the N spectra that best represent the observation with their corresponding EMDs, and N combinations of magnetic structures. Finally, we associate to each X-ray data set one best-matching EMD and, thus, one combination of magnetic structures, calculated as the average of the N selected EMDs and coverage fractions. In this way, we have indirectly derived the magnetic structures that contribute to the X-ray emission of a given observation.

\begin{figure*}[!htbp]
\centering
\subfloat{\includegraphics[width=0.45\textwidth]{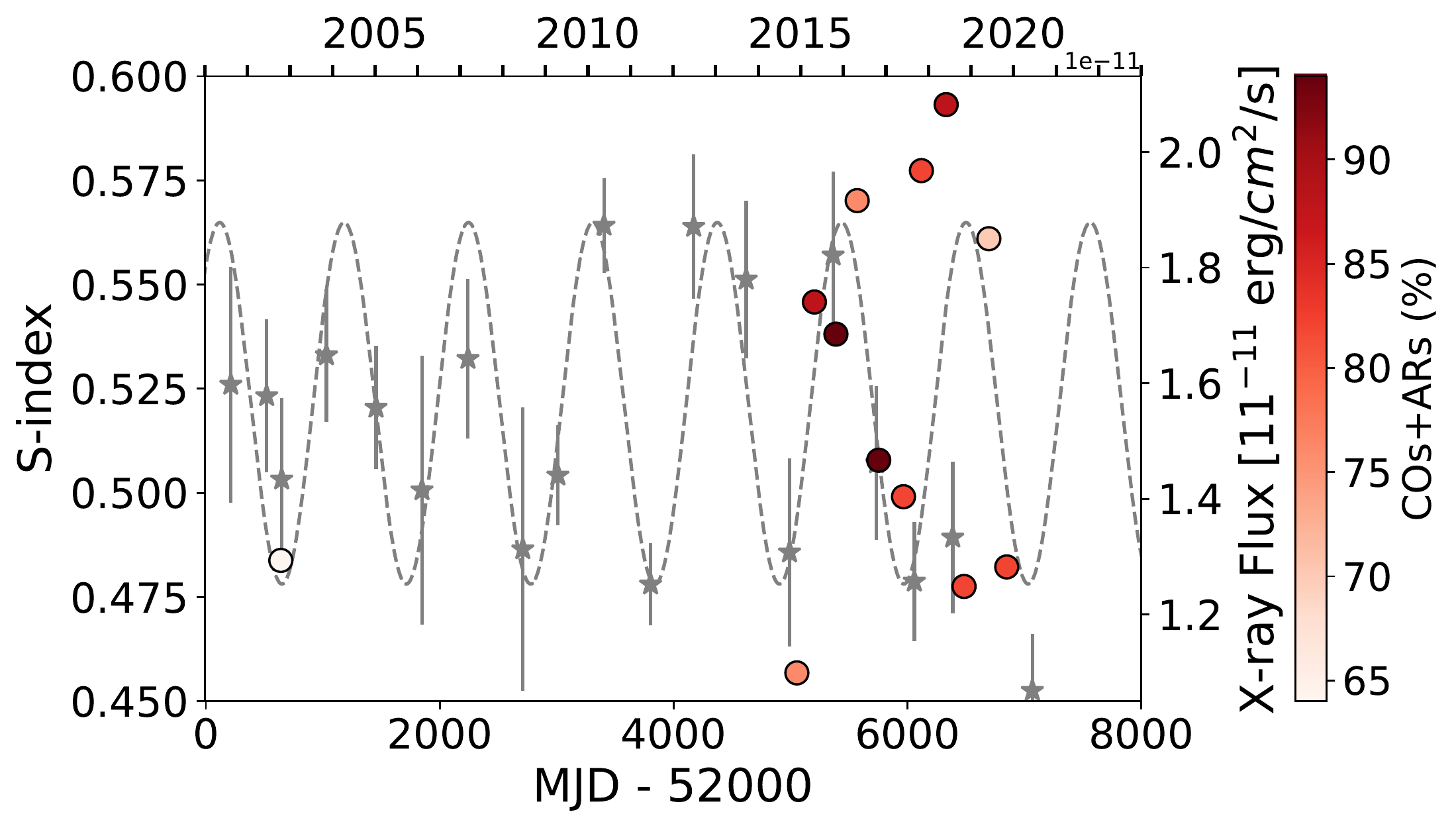}}
\subfloat{\includegraphics[width=0.45\textwidth]{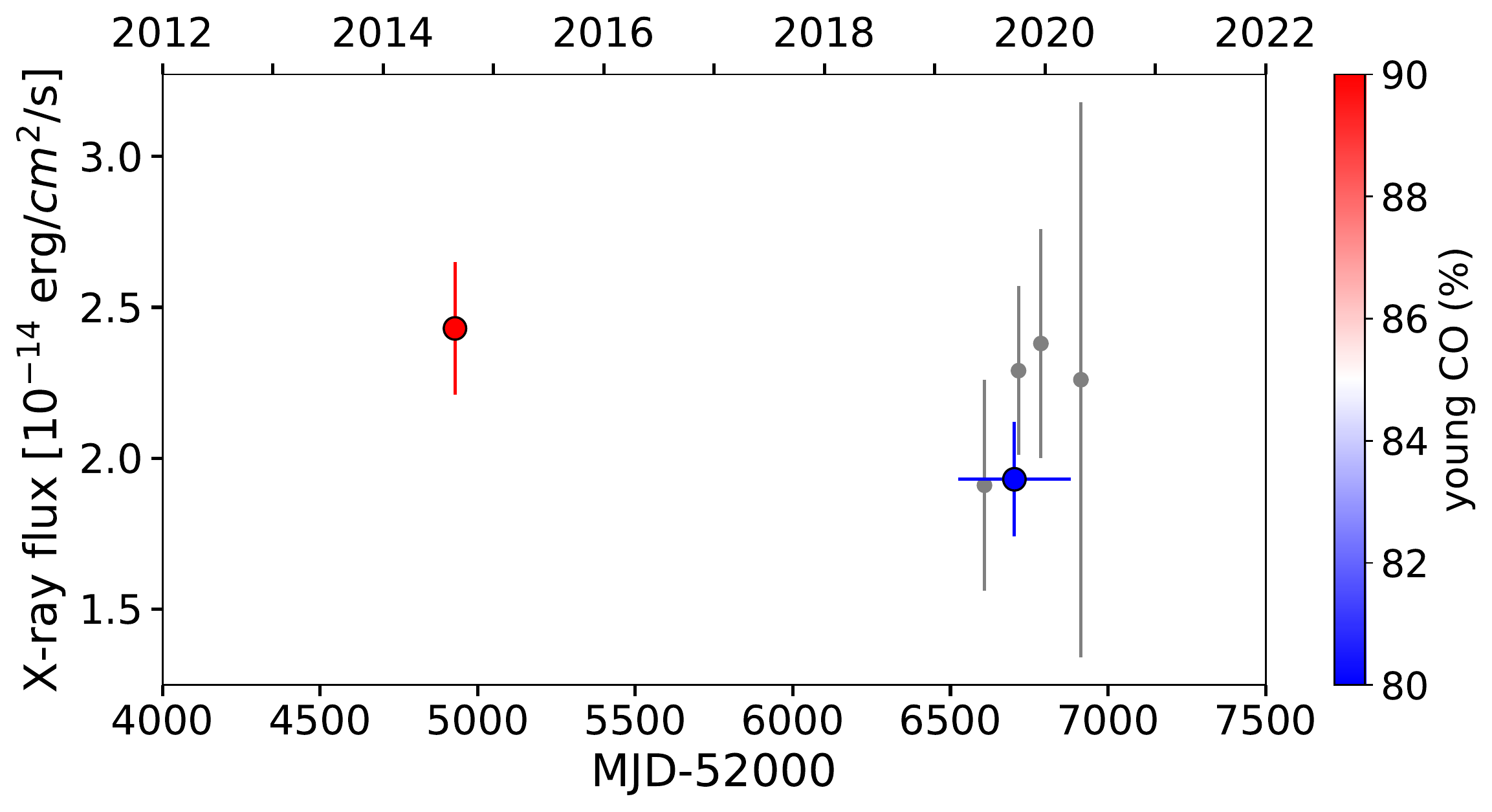}}
\caption{\textit{Left}: long-term lightcurve of $\epsilon$~Eri. The gray asterisks are the binned chromospheric S-index and the dotted line is the chromospheric cycle. The X-ray fluxes are plotted as colored circles. The color code refers to the different percentage of magnetic structures found from the simulations. \textit{Right}: long-term lightcurve of Kepler~63. The X-ray fluxes of the individual observations are plotted in gray. The colored dots are the two epochs to which the simulations were applied. }
\label{fig:lc}
\end{figure*}

\section{The coronae of young solar-like stars}
\label{sec:results}
We simulated the coronae of $\epsilon$~Eri and Kepler~63 in terms of combinations of solar magnetic structures by applying the method presented in \autoref{sec:sim}. 

In \autoref{sec:eeri} and in \autoref{sec:kep63} we summarize the main results obtained for the two stars, respectively. 

\subsection{$\epsilon$ Eri}
\label{sec:eeri}
Our sample of X-ray spectra consists of 12 \textit{XMM-Newton} observations of $\epsilon$~Eri \footnote{The X-ray monitoring campaign is still on going. Here, we present the X-ray observations of $\epsilon$~Eri, published by \citep{2020A&A...636A..49C} and spanning from 2003 to 2018, plus three new observations acquired in 2019 and at the beginning of 2020.} (Proposal ID: 074801; PI: France K.; Proposal IDs: 076049, 078024, 080116, 082007, 084345; PI: Stelzer B.).

The X-ray (EPIC/pn) spectra were analyzed with \texttt{xspec}, employing a 3-T APEC model. From the spectral fitting, we obtained the X-ray fluxes in the energy band $0.2-2$~keV and we constructed the long-term lightcurve of $\epsilon$~Eri, plotted in \autoref{fig:lc}on the left. Here, the colored dots are the X-ray fluxes, whereas the gray asterisks are the S-index measurements \citep{2020A&A...636A..49C}, binned across  time. The gray dotted line is the sinusoidal fit to the chromospheric data, that results in a period of $2.95$~yr. The variation of the X-ray flux of $\epsilon$~Eri results compatible with its chromospheric cycle.

From our X-ray monitoring, we found an average X-ray luminosity $\log L \ [erg/s] \sim 28.3$ (and a surface X-ray flux $\log F_{\rm X, s} \ [erg \ s^{-1} \ cm^{-2}] \sim 5.8 $) and an EM-averaged temperature $\log T \ [K] \sim 6.65$.
We applied our simulations starting from a grid of EMDs composed by ARs, COs and FLs. In particular, we assumed that the ARs cover $40\%$ of the corona of $\epsilon$~Eri and the COs can vary between $6$ and $60\%$. We also added a contribution of FLs to the EMD, testing: 1) models  in which the contribution of FLs is calculated by considering a sample of flares ranging from weak (class C) to bright (class X) ones and assuming their energy frequency distribution to be that in the solar corona (e.g., \citealt{2013NatPh...9..465W}; in the following MOD\_SOLAR); 2) models where only flares of class M, time-averaged over their evolution, contribute to the EMDs (in the following MOD\_FLM); 3) models where the flares of class M are considered only during their decay phase  (in the following MOD\_DECAY).

We extracted the corresponding X-ray spectra for each EMDs model using the \textit{XMM-Newton} response matrix and fitted with a 3-T APEC model. Then, the synthetic spectra were compared to the actual X-ray observations of $\epsilon$~Eri. The first two types of EMDs (that include the FL distributions MOD\_SOLAR and MOD\_FLM) are not able to properly reproduce the coronal temperatures observed on $\epsilon$~Eri and thus we considered only the third case (MOD\_DECAY), allowing the FLs to vary from $0$ to $10\%$ within the percentage of COs.

\begin{figure*}
\centering
\subfloat{\includegraphics[width=0.35\textwidth]{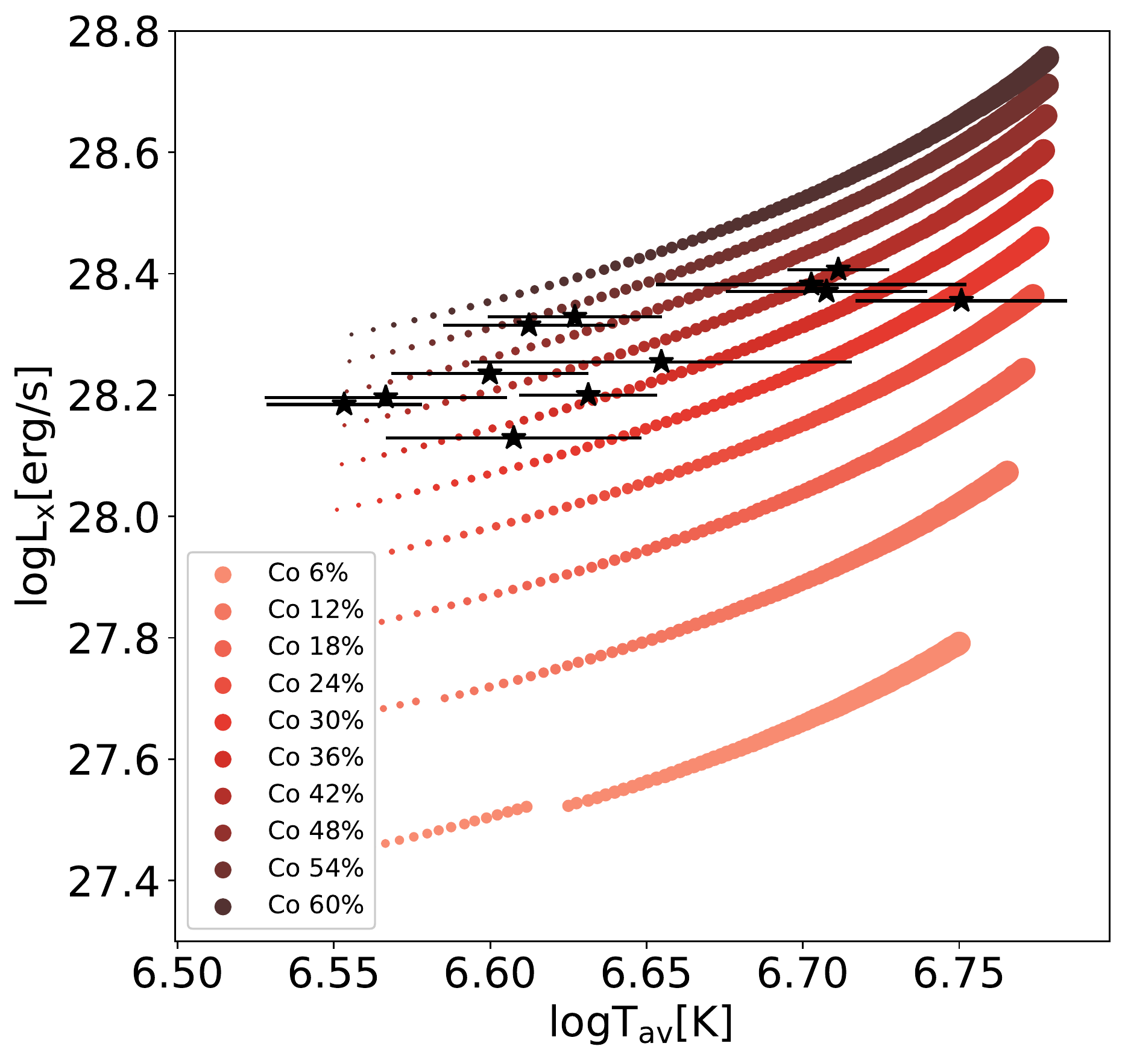}}
\subfloat{\includegraphics[width=0.35\textwidth]{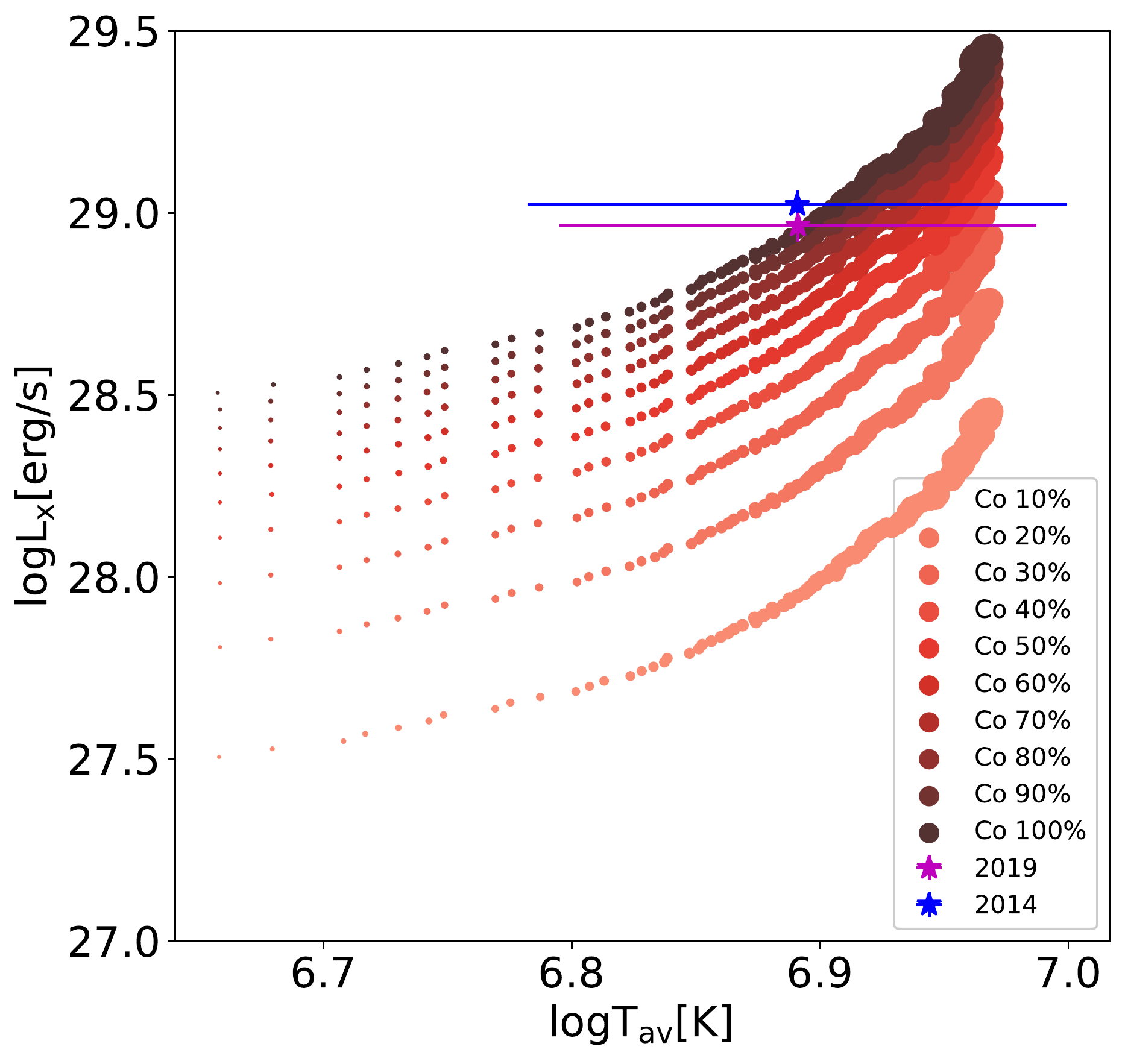}}
\caption{\normalsize{$L_{\rm X}-T_{\rm av}$ relation. On the left: X-ray observations of $\epsilon$~Eri (asterisks) and synthetic spectra (colored dots) obtained for ARs ($40\%$), COs ($6-60\%$) and FLs (MOD\_DECAY; $0-10\%$). On the right: X-ray observations of Kepler~63 (asterisks) and synthetic spectra (colored dots) obtained for young COs ($10-100\%$) and FLs (MOD\_FLM; $0-15\%$).}}
\label{fig:LxTav}
\end{figure*}

In the left panel of \autoref{fig:LxTav}, we show the $L_{\rm X}$-$T_{\rm av}$ relation calculated from the grid of the synthetic spectra (colored dots) and compared to that calculated for $\epsilon$~Eri (black asterisks). We notice that an increase in the filling factor of COs (changing color of the dots) leads to an increase of the X-ray luminosity. Conversely, an increase in the filling factor of FLs (increasing size of the dots) leads to an increase of the average temperature.

The left panel of \autoref{fig:LxTav} shows that the chosen combination of magnetic structures, i.e. $40\%$ of ARs, a range of COs ($6-60\%$) and FLs distribution of the type MOD\_DECAY, reproduces the X-ray luminosity and the average temperature of the actual X-ray spectra quite well. 

By applying the selection criterion from \autoref{sec:sim}, we found the best-matching EMDs to each observation of $\epsilon$~Eri, and therefore the corresponding percentage of magnetic structures. In the left panel of \autoref{fig:lc}, the sum of the coverage fraction of ARs and COs associated to each observation is indicated by the color bar. Our study, thus, suggests for $\epsilon$~Eri a coverage with magnetic structures that goes from $65\%$ to $95\%$ throughout the X-ray cycle.

\subsection{Kepler~63}
\label{sec:kep63}
The X-ray monitoring campaign of Kepler~63 comprises four X-ray observations, acquired between 2019 and 2020  (Proposal ID: 084123; PI: Coffaro M.). An additional archival observation was present in the \textit{XMM-Newton} archive, dating back to 2014 (Proposal ID: 074346; PI: Schmitt J.).  
The low photon statistics of the 2019-2020 observations allowed us to fit the spectra only with a 1-T APEC spectral model (Coffaro et al., 2021, in prep.). An isothermal model does not allow to decompose the EMD into different contributions and, thus, infer the coverage of ARs, COs, and FLs. Therefore, we simultaneously analyzed all observations acquired in 2019. With the higher statistics a 2-T APEC model can be used, and the same model is also applied to the longer exposure from 2014. Thus, we applied our method only to the observation of 2014 and to the observations of 2019, simultaneously fitted \footnote{The observation of 2020 is excluded, as the poor signal-to-noise ratio does not allow a fit, not even with a 1-T APEC spectral model.}. 

From the spectral analysis  of Kepler~63 we obtained a mean X-ray luminosity $\log L \ [erg/s] \sim 29$ (and a surface X-ray flux $\log F_{\rm X, s} \ [erg \ s^{-1} \ cm^{-2}] \sim 6.32$) and an average temperature $\log T \ [K] \sim 6.9$. 

For the application of our method, we excluded the ARs from the total EMD. This is motivated by the fact that even by covering the entire stellar corona with COs, the synthetic spectra predict X-ray luminosities that are lower than those observed on Kepler~63. Since the EMD per unit surface area of ARs is modest with respect to that of COs (\autoref{fig:arvsco}), assuming a partial coverage with ARs (and thus reducing the coverage with COs) would make the discrepancy between predicted and observed luminosities even higher. 
We found that the high $L_{\rm X}$ of Kepler~63 can be reproduced when newborn COs are considered. \cite{2004A&A...424..677O} first noticed that COs show a higher surface brightness at the beginning of their lives.  
We, therefore, built a total EMD composed of only young COs, varying from $10\%$ to $100\%$ on the coronal surface of Kepler~63. We also included the FLs, with  $<15\%$ of the coverage of COs, in order to properly reproduce the average temperature. We tested two different cases for the FLs contribution to the EMD: MOD\_SOLAR and MOD\_FLM (see \autoref{sec:eeri}).
We found that the MOD\_FLM is the one that better reproduces the observed $L_{\rm X}$ and $T_{\rm av}$ of Kepler~63.

We then analyzed the synthetic spectra with a 2-T APEC model. In the right panel of \autoref{fig:LxTav}, the $L_{\rm X}-T_{\rm av}$ relation obtained from the synthetic grid (colored dots) is shown in comparison with the one obtained for the two reduced epochs of Kepler~63 (blue and magenta asterisks). We notice that the synthetic grid reproduces the observations of Kepler~63 at the limit, assuming a full coverage of the stellar surface with young COs. There are no coronal structures observed on the Sun whose surface brightness can reach higher X-ray luminosity, nor can we add a higher number of structures as we reached the maximum  magnetic filling factor. Thus, the grid in \autoref{fig:LxTav} is the best choice for modeling the X-ray observations of Kepler~63 with solar coronal structures. Therefore we proceeded on the selection criterion. In the left panel of \autoref{fig:lc}, the coverage fraction with magnetic structures that is found from the quantitative comparison is reported in the color bar. We note that the two epochs of Kepler~63 data yield a CO filling factor of $80\%$ and $90\%$ in the corona.  

\section{Discussion and conclusions}
\label{sec:conc}
We presented a method that allows us to model the corona of stars in terms of solar magnetic structures and indirectly quantify these structures that are responsible for the X-ray activity of a star. Specifically, we characterized the X-ray emission of the two young solar-like stars $\epsilon$~Eri and Kepler 63. 

\begin{figure}[!htbp]
\centering
\includegraphics[width=0.35\textwidth]{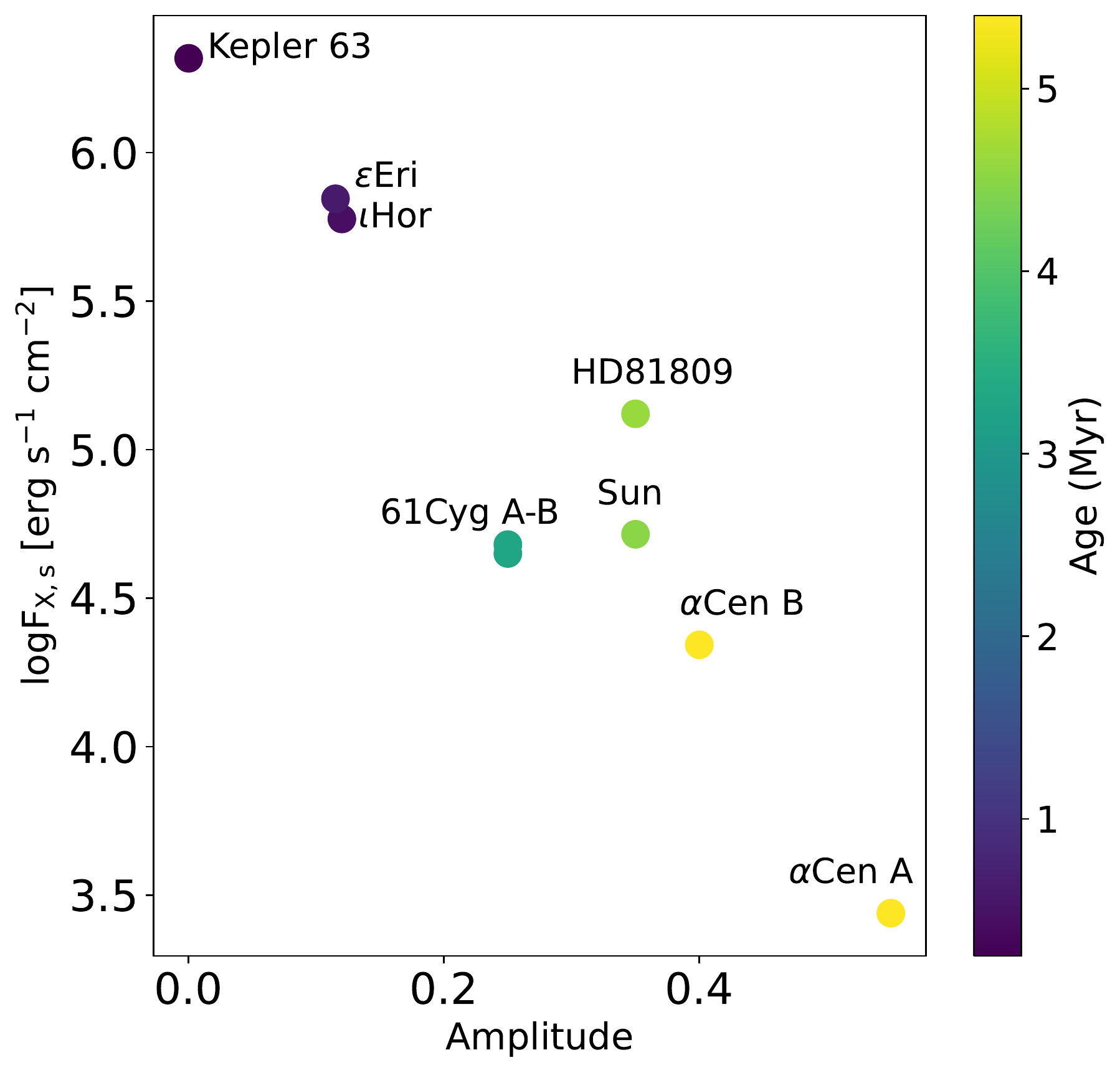}
\caption{Surface X-ray flux as a function of the cycle amplitude for all stars with known X-ray cycles. The color bar denotes the ages of the stars.}
\label{fig:amp}
\end{figure}

$\epsilon$~Eri displays a coronal cycle lasting $2.95$~yr (\autoref{fig:lc} on the left) and it is the youngest star so far known to have an X-ray cycle. Moreover, it also shows a remarkably small amplitude of the periodic X-ray variation compared to all other stars with known coronal cycles. This is shown in \autoref{fig:amp}, where the color bar refers to the ages of the stars\footnote{Kepler~63 is also included, even if \textit{XMM-Newton} did not revealed a coronal cycle.}. 
We see that overall as the surface X-ray flux decreases (and the age increases), the amplitudes of the coronal cycles increase. 
Our novel method to analyse the X-ray observations provides a physical explanation for this result.

The models point to a high coverage fraction with magnetic regions, that ranges from $65\%$ to $95\%$ of the coronal surface throughout the X-ray cycle.
This finding explains the high surface X-ray flux of $\epsilon$~Eri, as the result of a high magnetic filling factor in the corona, and therefore the small amplitude of its X-ray cycle: since the corona is almost full with magnetic structures, there is a lack of space for enhancing the covering fraction throughout the cycle and the X-ray luminosity can not strongly vary. Considering the well-known relation between $F_{\rm X}$ and stellar age, we conjecture that $\epsilon$~Eri sets a lower age limit for an X-ray cycle ($\sim 400$~Myr). Younger solar-like stars are expected to show a basal surface coverage with magnetic structures higher than what we have deduced for the case of $\epsilon$~Eri.

Our analysis based on solar magnetic structures well reproduces the coronal structure of $\epsilon$~Eri when we assume that the corona is covered by ARs, COs and FLs. We find a good match between actual X-ray observations of $\epsilon$~Eri and the model only if mid-energetic flares (M-class) in the decay phases are included in our study. This suggests that the flaring events that take place in $\epsilon$~Eri last longer than their solar counterparts, likely because of a lower coronal metallicity that makes radiative losses less efficient \citep{2020A&A...636A..49C}.

To further investigate coronal cycles in young solar-like stars, we studied an even younger target than $\epsilon$~Eri, Kepler~63 ($\sim 200$~Myr). The X-ray monitoring comprised five observations, and (from the spectral analysis) we do not see X-ray variability (\autoref{fig:lc} on the right). 

We applied our method reducing the five X-ray observations of the corona of Kepler~63 into only two observations. 
The models can fit the X-ray observations of Kepler~63 when we include only newborn COs, as such structures have a high surface brightness and thus are more suitable for reproducing the high X-ray luminosity of Kepler~63. Moreover, similarly to $\epsilon$~Eri, we also need a flaring component, and in particular mid-energetic flares (M-class), for reconstructing the temperatures of the corona. 

We found from the simulations that both \textit{XMM-Newton} data sets analysed for Kepler 63 show a high filling factor ($80-90\%$) with magnetic structures, even higher than those inferred for $\epsilon$~Eri. This result strengthens our hypothesis formulated above: in young stars cyclic X-ray variability is inhibited by a massive presence of coronal structures. 

In conclusion, with our new method of analysing stellar X-ray spectra with the help of solar observations, we indirectly gain information on the nature of the magnetic structures populating the stellar corona. Its application has shown that the corona of young solar-like stars can be modelled with types of magnetic structures that are observed on the Sun, but are not the ones that typically characterizes the solar X-ray cycle. This is not surprising as the present Sun is a more evolved star than the ones studied in this work, and its network of magnetic regions may have a different structure with respect to that present during its early life.
\section*{Acknowledgments}
MC acknowledges funding by the
Bundesministerium f\"ur Wirtschaft und Energie through the Deutsches Zentrum
f\"ur Luft- und Raumfahrt e.V. (DLR) under grant number FKZ 50 OR 2008.
S.O. acknowledges financial contribution from the agreement ASI-INAF n.2018-16-HH.0.
\bibliography{esac2021_MC}

\end{document}